\PassOptionsToPackage{table,xcdraw}{xcolor}
\documentclass[sigconf,dvipsnames,nonacm]{acmart}
\usepackage[utf8]{inputenc}
\usepackage[T1]{fontenc}
\usepackage{amsmath}
\usepackage{amsfonts}

\usepackage{graphicx}
\usepackage{xspace}
\usepackage{xcolor}

\usepackage{listings}
\usepackage{newfloat}
\usepackage{caption}
\usepackage{subcaption}

\usepackage{booktabs}
\usepackage{tabularx}
\usepackage{multicol}
\usepackage{multirow}
\usepackage{array}
\usepackage{hyperref}
\hypersetup{
	colorlinks=true,
	citecolor=blue,
	citebordercolor=green,
}

\graphicspath{{./figures/}}

\renewcommand{\paragraph}[1]{\vspace{0.5em}\noindent\textbf{#1.}}
\renewcommand{\subparagraph}[1]{\vspace{0.5em}\indent\textit{#1.}}

\newcounter{rqcounter}
\newcommand{\question}[1]{%
	\vspace{.5em}%
	\stepcounter{rqcounter}%
	\par\leftskip0cm\relax\rightskip0cm\relax%
	\textbf{RQ\therqcounter:\xspace#1}%
	\par\leftskip0cm\relax\rightskip0cm\relax%
	\vspace{.5em}%
	}

\newcounter{hypcounter}
\newcommand{\hyp}[1]{%
	\vspace{.5em}%
	\stepcounter{hypcounter}%
	\par\leftskip0cm\relax\rightskip0cm\relax%
	\textit{Hypothesis \textbf{H\thehypcounter}:\xspace#1}%
	\par\leftskip0cm\relax\rightskip0cm\relax%
	\vspace{.5em}%
}

\newcommand{\qt}[2]{``\textit{#2}''---\textsc{[P#1]}}

\newcounter{emailcounter}
\renewcommand{\theemailcounter}{\arabic{emailcounter}}

\usepackage{pifont}
\let\oldding\ding%
\renewcommand{\ding}[2][1]{\scalebox{#1}{\oldding{#2}}}
\newcommand{\one}{\ding[1.2]{192}\xspace}
\newcommand{\two}{\ding[1.2]{193}\xspace}
\newcommand{\three}{\ding[1.2]{194}\xspace}
\newcommand{\four}{\ding[1.2]{195}\xspace}
\newcommand{\five}{\ding[1.2]{196}\xspace}
\newcommand{\six}{\ding[1.2]{197}\xspace}
\newcommand{\seven}{\ding[1.2]{198}\xspace}
\newcommand{\eight}{\ding[1.2]{199}\xspace}

\newcolumntype{L}{>{\raggedright\let\newline\\\arraybackslash\hspace{0pt}}X}

\newcommand{\numusers}{4,554\xspace}

\newcommand{\minage}{18\xspace}
\newcommand{\maxage}{78\xspace}

\newcommand{\smallerGroup}{480\xspace}
\newcommand{\largerGroup}{534\xspace}

\title{Content, Nudges and Incentives: A Study on the Effectiveness and Perception of Embedded Phishing Training}

\author{Daniele Lain}
\affiliation{%
  \institution{ETH Zurich}
  \department{Department of Computer Science}
  \city{Zurich}
  \country{Switzerland}
}
\author{Tarek Jost}
\affiliation{%
  \institution{ETH Zurich}
  \department{Department of Computer Science}
  \city{Zurich}
  \country{Switzerland}
}
\author{Sinisa Matetic}
\affiliation{%
  \institution{ETH Zurich}
  \department{Department of Computer Science}
  \city{Zurich}
  \country{Switzerland}
}
\author{Kari Kostiainen}
\affiliation{%
  \institution{ETH Zurich}
  \department{Department of Computer Science}
  \city{Zurich}
  \country{Switzerland}
}
\author{Srdjan Capkun}
\affiliation{%
  \institution{ETH Zurich}
  \department{Department of Computer Science}
  \city{Zurich}
  \country{Switzerland}
}

\acmConference{}{}{}
\acmDOI{}
\acmISBN{}
\setcopyright{none}

\let\svthefootnote\thefootnote
\newcommand\freefootnote[1]{%
  \let\thefootnote\relax%
  \footnotetext{#1}%
  \let\thefootnote\svthefootnote%
}

\begin{document}

\begin{abstract}
	A common form of phishing training in organizations is the use of simulated phishing emails to test employees' susceptibility to phishing attacks, and the immediate delivery of training material to those who fail the test.
This widespread practice is dubbed \textit{embedded training}; however, its effectiveness in decreasing the likelihood of employees falling for phishing again in the future is questioned by the contradictory findings of several recent field studies.

We investigate embedded phishing training in three aspects.
First, we observe that the practice incorporates different components---knowledge gains from its content, nudges and reminders from the test itself, and the deterrent effect of potential consequences---our goal is to study which ones are more effective, if any.
Second, we explore two potential improvements to training, namely its timing and the use of incentives.
Third, we analyze employees' reception and perception of the practice.
For this, we conducted a large-scale mixed-methods (quantitative and qualitative) study on the employees of a partner company.

Our study contributes several novel findings on the training practice: in particular, its effectiveness comes from its nudging effect, i.e., the periodic reminder of the threat rather than from its content, which is rarely consumed by employees due to lack of time and perceived usefulness.
Further, delaying training to ease time pressure is as effective as currently established practices, while rewards do not improve secure behavior.
Finally, some of our results support previous findings with increased ecological validity, e.g., that phishing is an attention problem, rather than a knowledge one, even for the most susceptible employees, and thus enforcing training does not help.
\end{abstract}

\maketitle

\pagestyle{plain}

\freefootnote{Extended version of the paper to appear in the proceedings of ACM CCS '24.}
\freefootnote{\url{https://doi.org/10.1145/3658644.3690348}}

\section{Introduction}
\label{sec:introduction}
Phishing attacks represent one of the main threats for organizations~\cite{ibmreport}.
This form of cybercrime consists of deceiving Internet users to perform some unsafe behavior that benefits the criminals, such as clicking on a malicious link and entering their credentials or downloading malware from websites or email attachments~\cite{aleroud2017phishing}.
In organizations, this often takes the form of emails~\cite{cloudian2021survey,ibmreport} where the adversaries pose as trustworthy entities such as reputable companies or colleagues.
These attacks are on the rise, favored by the increase of remote work and the COVID-19 pandemic~\cite{bitaab2020scam} and represent a constant source of financial loss for companies~\cite{phishingcost}.

To combat this threat, there are significant efforts from academia and industry to stop these attacks during all their stages~\cite{khonji2013phishing}, from filtering email~\cite{salloum2022systematic} to detecting malicious URLs~\cite{basnet2014learning} and websites~\cite{tang2021survey}.
However, the cat-and-mouse nature of phishing~\cite{oest2018inside} means that phishing eventually gets to the employees, who become the last line of defense.

Training employees to recognize phishing attacks is therefore a crucial part of an organization's security strategy~\cite{jampen2020don}.
While cybersecurity training and awareness can take many forms~\cite{wash2018provides}, a widespread industry practice is \textit{``embedded training''}~\cite{kumaraguru2010teaching}: the use of simulated phishing emails together with training material that is presented to employees who fail the test and fall for the phish.
This is based on the principle that learning next to one's mistake is an effective teaching strategy~\cite{kumaraguru2007protecting}; however, this practice has been under scrutiny due to unclear performance in field studies~\cite{caputo2013going,siadati2017measuring,lain2022phishing} and potential side effects on employees' morale~\cite{kirlappos2015fixing,hielscher2023employees,caputo2013going} and well-being~\cite{reeves2021encouraging}.

\paragraph{Our study}
In this paper, we investigate the practice of embedded phishing training in organizations in the following three aspects.

First, we investigate the \textit{effectiveness} of training: recent studies showed contradicting results on the effectiveness of the practice in improving employees' future behavior~\cite{caputo2013going,siadati2017measuring,lain2022phishing}.
Further, we observe that there are different components contributing to its effect, such as the gains in knowledge coming from its content, the nudging and reminder factors caused by the tests, and the deterrent effect of potential consequences.
Which of these components are responsible for the effects of training, if any, is unclear~\cite{marshall2023exploring,sutter2022avoiding}.

Second, we investigate two potential \textit{improvements} to the practice.
The first improvement we investigate regards when to deliver training: since time pressure is one of the main reasons for people to fall for phishing~\cite{butavicius2022people}, we ask ourselves whether it is beneficial to delay serving training, instead of doing it immediately after an employee fails a phishing test, as it is commonly done in organizations~\cite{volkamer2020analysing}.
The other improvement we investigate is the use of rewards in phishing reporting: many companies attach some form of \textit{``sticks or carrots''} to their phishing training practices~\cite{blythe2020human}, however, negative elements such as punishments are more common than positive incentives~\cite{blythe2020human}, which we instead study in this paper.

Third, we study employees' \textit{reception} and perception of phishing training.
The practice is perceived by security professionals as a potential source of conflict with employees~\cite{hielscher2023employees,kirlappos2015fixing}; further, employees being confused by the simulations and the training is hypothesized~\cite{lain2022phishing} as one of the reasons for recent paradoxical results, such as employees performing better without training~\cite{caputo2013going,lain2022phishing}.

We study these aspects in a large-scale mixed-methods study with a partner company.
We implement these changes and different experimental conditions within the company's established embedded phishing training and testing.
We analyze the performance of \numusers participants from the company's employee base in response to three simulated phishing emails sent over 6 weeks.
These emails were received as part of the participants' normal workflow; further ecological validity comes from a wide diversity of participants in terms of roles, responsibilities, seniority, and types of jobs at the partner organization.
Besides the measurement of employees' performance in the phishing tests, we interviewed 25 participants to gather qualitative insights on their perception of the practice.

Our study provides two main contributions: first, we provide several novel findings regarding the effectiveness and reception of embedded phishing training.
Among our main findings, we found that the effectiveness of the practice derives from its nudging effect, i.e., the periodic reminder of the threats of phishing and to remain vigilant, rather than the content of the training, which we observe is deemed as not helpful even by the most susceptible participants.
This is of particular interest given that the common perception of the practice is that it works by increasing knowledge~\cite{zhuo2023sok}.
Moreover, we observe that the training practice is a potential source of misunderstandings and overconfidence in employees.
We further found that delaying training to a later time when there is less time pressure is as effective as the ``immediate'' embedded one, and that introducing rewards is not beneficial to the practice.
Second, we support previous findings regarding training, but on a more diverse and larger population: for example, enforcing training on employees more susceptible to falling for phishing is not beneficial~\cite{gordon2019evaluation}, and susceptibility itself is more related to lack of attention rather than lack of domain knowledge~\cite{canham2023repeat}.

To summarize, our paper makes the following contributions:
\begin{itemize}
    \item A large-scale mixed-methods (quantitative and qualitative) study on the effectiveness and perception of embedded phishing training in organizations.
    \item Novel findings regarding which components of embedded phishing training are responsible for its effectiveness (Section~\ref{sec:results_effectiveness}).
    \item New results on two potential improvements to the practice: timing of training, and the use of rewards (Section~\ref{sec:results_improvements}).
    \item Qualitative insights on employees' reception and perception of the practice (Section~\ref{sec:results_reception}).
\end{itemize}

\paragraph{Outline}
The rest of the paper is organized as follows.
In Section~\ref{sec:background}, we provide background on the practice of embedded phishing training and related work studying the effectiveness and reception of such practice.
In Section~\ref{sec:rqs}, we present our research questions and main findings, and Section~\ref{sec:expsetup} describes the experimental setup of our study.
In Section~\ref{sec:results_effectiveness}, we report the results of our study regarding the effectiveness of training and its components; in Section~\ref{sec:results_improvements}, the studied improvements to the practice; and in Section~\ref{sec:results_reception} its reception and perception by employees.
In Section~\ref{sec:discussion}, we discuss the validity of our study.
Section~\ref{sec:conclusion} concludes the paper.

\section{Background and Related Work}
\label{sec:background}
\paragraph{Phishing Training}
One of the main tools for organizations to fight phishing is training their employees to recognize such attacks, e.g., how to identify suspicious emails~\cite{harrison2016individual}, URLs~\cite{sheng2007anti} and attachments~\cite{wen2019hack}, and websites~\cite{alnajim2009anti}, which are the most common vectors.
This training can take the form of dedicated classroom sessions~\cite{marshall2023exploring}, online resources~\cite{bian2009evaluation}, games~\cite{sheng2007anti} or exercises~\cite{kumaraguru2010teaching}, with wide consensus that it should preferably be some form of active learning task~\cite{marshall2023exploring}.

To measure the resilience of employees to phishing attacks and how it changes over time, e.g., as a consequence of training, organizations commonly use phishing simulations: sending fake phishing emails to their employees that try to entice them into performing something unsafe, e.g., opening an attachment or entering their credentials on a website, and measuring how many fall for them~\cite{kumaraguru2007protecting}.

\paragraph{Embedded Training}
One of the most widespread forms of phishing training is \textit{embedded training}~\cite{kumaraguru2007protecting}, which moves the training from a dedicated session to the moment of the mistake in a phishing test.
This form of training, based on the idea that learning next to one's mistake is an effective teaching strategy~\cite{kumaraguru2010teaching}, employs phishing simulations to the employees who, if they fall for the test by, e.g., clicking on a suspicious link, are immediately redirected to training material.
This training is usually delivered in the form of a webpage with training material that explains the exercise, the mistake, and how to avoid it in the future, potentially presenting further information with different media such as videos, games, quizzes, or e-learning environments~\cite{marshall2023exploring}.

\paragraph{Training Effectiveness}
Numerous studies on the practice of embedded phishing training have contradicting findings on its efficacy~\cite{zhuo2023sok,marshall2023exploring}.
Recent literature reviews of the area indicate that these works are hard to compare due to different conditions and studied populations, potentially leading to disagreements in results, i.e., whether click and fall rates for phishing simulations actually decrease~\cite{sutter2022avoiding,jansson2013phishing} or stay unchanged~\cite{gordon2019evaluation} or even increase~\cite{lain2022phishing,caputo2013going} as a consequence of training.

It is therefore unclear which elements of training, if any, improve performance on future phishing tests, e.g., recent studies observing improvements found that most participants did not actually read the training material~\cite{sutter2022avoiding,jansson2013phishing}, and large meta-analyses report inconclusive results on the effectiveness of knowledge gains~\cite{zhuo2023sok,marshall2023exploring}.
While this practice is widespread~\cite{blythe2020human,hielscher2023employees}, studies are starting to investigate its costs and associated challenges~\cite{brunken2023properly} and, since organizations are mostly looking for simple metrics from the practice rather than proper education~\cite{hielscher2023employees}, questioning whether the practice's unclear benefits justify the increased friction with employees~\cite{caputo2013going} and its added mistrust~\cite{kirlappos2015fixing,blythe2020human} and fatigue~\cite{reeves2021encouraging}.

Additionally, phishing tests in organizations often come attached with the (threat of) consequences. Some of these can be explicit and known by employees (e.g., losing Internet access until a training program is completed, to more serious consequences up to risking one's job~\cite{blythe2020human}), but the absence of any explicit policy can anyway be associated with a general fear of repercussion from management.
This is a common setting in many organizations~\cite{herath2009encouraging}, despite attaching deterrents to phishing training has been shown to be more detrimental than beneficial~\cite{brunken2023properly} and can further increase the \textit{cyber fatigue} caused by phishing tests~\cite{reeves2021encouraging}.

To investigate these contradicting results, work is being done to understand who in organizations~\cite{lain2022phishing} and the general public~\cite{sheng2010falls} falls for phishing and why~\cite{butavicius2022people,alsharnouby2015phishing,sarno2022so}, why ``repeat clickers'' apparently resilient to any form of training exist across different studies and organizations~\cite{canham2019enduring}, and whether it is a matter of personality~\cite{pattinson2012some} or lack of knowledge~\cite{canham2023repeat} or awareness~\cite{jansson2013phishing}.

\section{Research Questions \& Main Findings}
\label{sec:rqs}
\subsection{Training Effectiveness}

The most realistic way to assess the real impact and effects of employee phishing training is by studying it in large organizations.
However, most organizations are already running some form of training program, and it is therefore likely that many elements potentially contributing to employee performance are already coupled, e.g., the training content itself, the \textit{nudging} effect of failing a phishing test that reminds employees to be more attentive, and potential consequences if too many tests are failed.
Consequently, the practice of embedded phishing training is usually studied as a whole~\cite{caputo2013going,gordon2019evaluation,jampen2020don,zhuo2023sok,sutter2022avoiding} and through a single metric---evolution over time of click and fall rates on simulated phishing---that we observe does not solely capture the effects of learning from training material.
For example, this metric is also influenced by the nudging effect caused by seeing the training material after failing a phishing test, which can potentially lead to a short-lived improvement that reflects the employee being more vigilant rather than an increase in knowledge or a change in behavior.
Therefore, we ask ourselves:

\question{Which components of embedded phishing training lead to improvements in future phishing detection?}

To study this, we separate these different components, i.e., (i) training content, (ii) the reminder effect, and (iii) the deterrent effect of potential consequences, and understand how they might differently contribute to the metric of click and fall rates.
First, we investigate how state-of-the-art embedded training practices impact employees' future performance in detecting phishing emails:
\hyp{Embedded training improves future phishing detection.}

\noindent And also investigate the impact of simple deterrents in case of failing a phishing test:
\hyp{Deterrents improve future phishing detection.}

Then, we compare these two settings to understand which underlying elements contribute to employees' improvements: while both training and deterrents provide a nudge to employees, only training provides content to help employees understand what to look for in phishing emails.
Thus, we study the usefulness of the content of training material as follows:
\hyp{Embedded phishing training is more effective than deterrents in improving future phishing detection.}

Finally, to properly assess the teaching factor of the content of training material, we need to consider whether it is actually consumed.
However, some companies do not actively check this and only use phishing tests to assess the organization's vulnerability to threats~\cite{hielscher2023employees}---this is the case for our partner company who does not enforce reading the embedded training material, which is left as voluntary.
We cannot explore this aspect fully without radical changes in the processes of the organization we collaborate with.
However, we can study whether the enforcement of going through the training material is beneficial for a population of special interest: users more susceptible to falling for phishing tests.
We formulate the following hypothesis:
\hyp{Enforced training is more effective than voluntary training and no training for more susceptible participants.}

We note that besides investigating its overall contribution to employees' alertness to phishing, studying deterrents might not seem useful, as there is evidence that attaching repercussions to it is more detrimental than beneficial~\cite{kirlappos2015fixing,hielscher2023employees,herath2009encouraging}.
However, it is a common practice in organizations~\cite{marshall2023exploring}; further, besides explicit repercussions, there is an implicit threat of consequences and negative feelings associated with making mistakes, thus worth studying.

\subsection{Training Improvements}

We observe that it is unclear when the best time to provide training material to the employees is.
On one side, the practice of embedded training of having a ``teachable moment'' following a mistake is considered an effective teaching practice~\cite{kumaraguru2010teaching}; however, phishing susceptibility is often associated with high workload and time pressure~\cite{butavicius2022people}---indeed, causing a sense of urgency is one of the most used strategies by phishers~\cite{williams2018exploring}.
We therefore ask ourselves whether such principle of serving the training material immediately after the mistake employed by embedded training represents the best moment to deliver training. 

As discussed, while many organizations attach deterrents to their phishing exercises~\cite{blythe2020human}, this practice is criticized in the literature~\cite{kirlappos2015fixing,herath2009encouraging}.
We observe that, instead, positive incentives to secure behaviors and to good performance in these exercises are less studied.
We ask ourselves whether the prospect of a reward to employees for correctly handling the training exercises can be beneficial.

From these two observations, we investigate the following improvements to the training practice: (i) we delay the delivery of training material to a later moment when the employee might be more receptive to training; and (ii) we introduce rewards for correctly reporting simulated phishing emails.
We ask ourselves whether these improvements are beneficial:

\question{Can we improve the effectiveness of phishing training by delaying content delivery or introducing rewards?}

\paragraph{Training timing}
We hypothesize that someone failing a phishing exercise might not have the immediate time to go through some learning material or to get a lecture on phishing.
Thus, we study whether it is helpful to delay informing the user that they failed a phishing test and only provide training material after a few hours:
\hyp{Delaying delivering training material is more effective than embedded training in improving employees' future phishing detection.}

\paragraph{Incentives}
We observe that the presence (or absence) of additional motives can influence employees' performance on phishing exercises and their motivation to go through training material~\cite{pfleeger2012leveraging}.
For example, incentives to report phishing might motivate employees who would not be inclined to do so by their personality~\cite{marin2023influence}, and nominated security ``champions'' can advise protect their peers~\cite{burda2020don}.
We investigate the impact on employees of rewarding them for reporting phishing emails both on their future performance on phishing exercises:
\hyp{Rewards improve future phishing detection.}

\noindent And also on their rates of reporting phishing emails:
\hyp{Rewards increase the rate of reported emails.}

\subsection{Training Reception}

It is important to study how phishing tests are received: employees' opinion of them can potentially influence their performance (up to the point of boycotting the disliked practice by making mistakes on purpose), and they are associated with negative feelings (feeling judged or shamed for failing a test~\cite{hielscher2023employees}) and stress~\cite{reeves2021encouraging,blythe2020human,hielscher2023employees}.
Further associated with employees' perception is their understanding of the nature of phishing tests: recent work observing poor efficacy of embedded phishing training~\cite{lain2022phishing} hypothesizes that the whole process might be misunderstood, e.g., employees could believe the training material is a company appliance protecting them from \textit{real} phishing, in turn causing overconfidence in the security of their corporate email.
Other hypotheses for this behavior are that the ``gamified'' aspect of embedded training might cause employees not to take it seriously, that phishing is not perceived as a real threat, or that the provided training material is simply not effective in those employees who are most vulnerable to phishing emails.

\question{Is the practice of embedded phishing training perceived as useful, well received, and understood?}

To answer this, we do not set explicit hypotheses but rather aim to gather qualitative data on (i) which aspects of the tests are perceived as more or less useful; (ii) feelings towards the practice; and (iii) understanding of the nature of phishing tests.

\subsection{Summary of Main Findings}

\paragraph{Training Effectiveness}
There are several studies investigating the performance of embedded phishing training in large organizations~\cite{caputo2013going,gordon2019evaluation,jampen2020don,zhuo2023sok,sutter2022avoiding,volkamer2020analysing}.
Our study presents similar findings to recent work that observes improvements in employees' future behavior after training~\cite{sutter2022avoiding,jansson2013phishing} compared to a control population.
Further, our study reinforces the empirical observation of these previous studies that employees improve despite not reading the training material~\cite{sutter2022avoiding,jansson2013phishing} with qualitative evidence: even the most vulnerable participants in our study did not feel that the training content presented new or useful information.
We further support on a larger scale with increased ecological validity the observations of a previous smaller pilot study that suggested that phishing susceptibility is more related to lack of attention rather than lack of knowledge~\cite{canham2023repeat}.

The novel contributions of our study regarding effectiveness are twofold.
First, by comparing the different components of training (\textit{H3}), we find that the nudging element of our deterrent is as effective as training material in preventing employees from falling for future phishing attempts (or even reduces the likelihood of clicking on future phishes), suggesting that the content of training material is not beneficial.
Second, we contribute an explanation to the paradoxical results of improvements despite not consuming the training content: the component of training that offers benefits to employees lies in its periodic reminder, rather than in the increase of knowledge.

Our data finally suggests that the enforcement of training for the most susceptible employees (\textit{H4}) does not lead to a significant improvement in their performance and could be potentially poorly welcomed.
This supports previous findings on a more specialized population~\cite{gordon2019evaluation} and extends them to a more diverse population.

\paragraph{Training Improvements}
Our study provides two novel insights into the timing of training.
First, we show that informing employees about the exercise and directing them to training material the day after the incident (\textit{H5}) seems to be as effective as ``immediate'' embedded training.
Second, we observe that time pressure is not only a contributing factor in falling for phishing (which was already studied~\cite{sarno2022so,harrison2016individual}) but also in not completing training: this represents a novel finding and suggests that the timing of training needs to be further studied---especially if it can ease the pressure on employees.

Regarding the adoption of positive incentives in the form of rewards for reporting phishing (\textit{H6}), our quantitative and qualitative data suggest that there is no observable benefit.
While crowdsourced phishing reporting appears to be an effective tool in organizations~\cite{burda2020don,lain2022phishing} and the motives and traits of phishing reporters are actively being studied~\cite{zhuo2023sok,burda2020don,marin2023influence}, our novel insight is of interest because many companies experiment with rewards in their phishing training and awareness campaigns~\cite{blythe2020human,herath2009encouraging} but without evidence of their effectiveness.

\paragraph{Training Reception}
We find that the practice can be misunderstood, as a part of our interviewed population failed to understand the nature of phishing tests and confused them with real attacks, and the training page by the attack being thwarted by the company's IT.
Further, training can make employees overconfident both in their abilities and in the fact that mistakes in phishing tests are without repercussions.
This is a novel insight, that confirms the hypotheses formulated but not yet tested by recent studies investigating why the training practice can lead to the paradoxical results of increasing future click and fall rates on simulated phishing emails~\cite{lain2022phishing}.

Contrary to previous studies reporting negative sentiments towards the practice~\cite{caputo2013going,hielscher2023employees}, we observe generally positive sentiments, albeit in a company where it has been in place for many years and where employees are now used to it. However, our volunteer-based sample of interviewed participants suggests caution.

\section{Experimental Setup}
\label{sec:expsetup}
\begin{figure*}[!th]
	\centering
	\includegraphics[width=\linewidth]{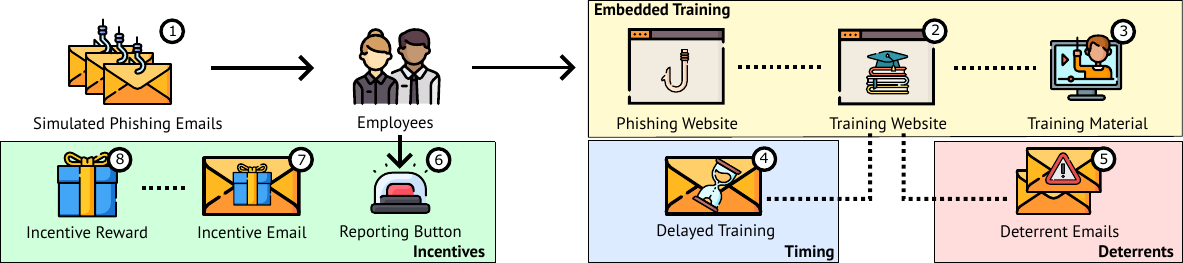}
	\caption{Experimental setup in the partner company infrastructure.}
	\label{fig:setup}
\end{figure*}

\subsection{Study Organization}
\label{sec:expsetup.organization}

\paragraph{Partner company}
We collaborated with a company that is active in different businesses in Switzerland and has around 60,000 employees with heterogeneous professions, training, education, and specialization levels.
As part of its cybersecurity measures, the company has different processes in place for phishing prevention and awareness.
In particular, employees get onboarding training on the risks of phishing when they join and yearly training refreshers. Further, the company routinely employs simulated phishing emails as a tool to assess their employees' vulnerability to this threat, and employees falling for one of these simulations are immediately redirected to an informational webpage regarding the failed test with an option to go through further self-training, i.e., \textit{embedded phishing training}.
These training and tests are provided by a specialized external company.

\paragraph{Infrastructure}
The experiment setup is depicted in Figure~\ref{fig:setup}.
Large-scale ecological studies like ours can only be conducted in large organizations, where it is likely that there are already established processes and tools for phishing prevention and training. 
Therefore, we designed our study to fall within the mechanisms already in place by introducing alternative settings and additional elements to answer our research questions. 
We now overview the different components of the infrastructure: the ones that were already in place and the ones we introduced for this study.

\subparagraph{Existing Infrastructure}
The company already regularly uses \textit{simulated phishing emails} (marked as~\one in Figure~\ref{fig:setup}) to assess their employees' resilience to phishing.
These emails usually contain a link to a (simulated) phishing website, that asks employees to do something unsafe, e.g., entering their corporate credentials. Employees that do so get immediately redirected to a training website~\two, developed by a vendor and hosted on an internal corporate website, that details to the employee that they failed a phishing test, information on the nature of the exercise, an explanation of how the phish could have been spotted, and instructions that suspicious emails should be always reported.
The page further directs to additional learning resources~\three developed either by the company or their vendor: an instructional video, an e-learning environment, and quizzes to further test one's knowledge.
Such design follows the state-of-the-art in embedded training~\cite{kumaraguru2010teaching}.
 
The company also provides its employees a button on their email client~\six allowing them to report phishing emails to IT.
The button is generally known to the employees, having been in place for two years before the start of the experiment

\subparagraph{Training Timing} 
To test training timing, we introduced a debriefing email~\four to send the morning after an employee failed a phishing test. 
This email with corporate design is sent by the partner company's IT and briefs the employee that they failed an exercise the previous day, inviting them to go through training by linking the embedded training webpage~\two. Employees had to go through delayed training within 7 days from the mistake and notice; this was enforced by the participant's supervisor.
We provide the content of this email in Appendix~\ref{sec:extra.study_emails}.

\subparagraph{Deterrents}
We deployed deterrents in the form of two emails~\five. 
The first one is sent after an employee failed a phishing test for the first time during the experiment and warns that failing another will require them to undergo mandatory training on the dangers of phishing.
The second one, sent after failing a second time, informs that the employee now has to undergo training within the next week and links the training webpage also served for embedded training~\two.
Similarly to the delayed training emails, these emails were sent the morning after an employee failed a phishing test by the partner company's IT administration.
However, they also included in copy the employee's manager, who was responsible for checking that the employee completed the training.
The content of these emails is also provided in Appendix~\ref{sec:extra.study_emails}.

We tied our choice of deterrent to enforcing employees going through the training material for two reasons: first, it is a light repercussion that is not putting participants at risk.
Second, we wanted to study the effects of training content when we are sure it is actually consumed---it was not possible in our partner company to enforce training for a large subset of employees, nor to accurately record data such as the time spent on the training website, so we opted to study enforcing reading the training material only for the subset of participants that compose their most susceptible ones, i.e., those that failed two phishing tests.

\subparagraph{Incentives}
To study the effect of incentives, we introduced them as part of the company's existing phishing reporting mechanism.
Choosing an active action such as reporting gives us the benefit of knowing for sure that the employee detected the phish---otherwise, if the employee does not click or fall for the test, we cannot know whether it was simply ignored or overlooked.
We recall that, prior to our study, employees already had a button on their email client to report phishing to IT~\six. 
We introduced an email sent at the start of the experiment~\seven that informed the recipient they would be rewarded with a prize if they correctly reported a phishing email, as part of a pilot program (reported in Appendix~\ref{sec:extra.study_emails}).
The prize~\eight, set by the company as a box of chocolate, was sent by IT in weekly batches.

\subsection{Study Participants}
\label{sec:expsetup.participants}

The company enrolled \numusers employees to participate in our study---henceforth \textit{participants}.
Selection happened at random from the whole company's employee base, comprised of both people ``in the field'' (from logistics to storefronts) and office roles such as accounting, technical, and managerial positions.
Participants were aged \minage to \maxage (see Figure~\ref{fig:age_groups}), and skewed to the age range 50 to 59, reflecting the composition of the company's employee base; 2,594 were male and 1,960 female.

A small subset of these participants took part in follow-up interviews.
Criteria for inclusion were based on performance during the study: we contacted both participants who never failed any phishing test and participants who failed at least two out of three to compare their different experiences.
Based on these criteria, the company contacted around 200 matching participants (loosely distributed among the different experimental groups we detail in Section~\ref{sec:expsetup.groups}) inviting them for a voluntary 20/30 minute interview with the researchers, without remuneration but during their work hours.
Of these, 25 participants accepted and took part in semi-structured interviews: 5 ``experts'' that did not fall for any phishing test, and 20 that failed at least two out of three.

\subsection{Experimental Groups}
\label{sec:expsetup.groups}

\begin{table}[t]
	\renewcommand{\arraystretch}{1.2}
	\centering
	\caption{Experimental groups for our study and their administered settings. We display the number of participants in each group and the number of participants that fell for at least one phishing test.}
	\rowcolors{2}{white}{gray!10}
	\begin{tabularx}{\linewidth}{@{}llXXX@{}}
		\toprule
		& Training type
		& Incentive / \newline Deterrent
		& Participants
		& $>0$ dangerous actions
		\\
		\midrule

		1 & \cellcolor{white} & Incentive & 483 & 110 \\ 
		2 & \cellcolor{white} & Deterrent & 522 & 143 \\ 
		3 & \multirow{-3}{*}{\cellcolor{white}Immediate} & No & 500 & 151 \\
		\midrule
		4 & \cellcolor{gray!10} & Incentive & 480 & 103 \\ 
		5 & \cellcolor{gray!10} & Deterrent & 516 & 135 \\ 
		6 & \multirow{-3}{*}{\cellcolor{gray!10}Delayed} & No & 534 & 146 \\
		\midrule
		7 & \cellcolor{white} & Incentive & 480 & 123 \\ 
		8 & \cellcolor{white} & Deterrent & 529 & 136 \\ 
		\textit{Control} & \multirow{-3}{*}{\cellcolor{white}None} & No & 510 & 127 \\

		\bottomrule
	\end{tabularx}
	\label{tab:groups}
\end{table}

Each study participant was randomly assigned to one of 9 different \textit{experimental groups} according to the different settings that were administered regarding training timing, incentives, and deterrents:

\begin{itemize}
	\item \textbf{Training timing}: after failing a phishing test, participants either received the pre-existing state-of-the-art ``immediate'' embedded training, our delayed training, or no training (as part of a control group).
	\item \textbf{Deterrents}: participants with the deterrent setting were sent the ``first and second strike'' emails we deployed after failing a phishing test; or no such emails otherwise.
	\item \textbf{Incentives}: participants were either promised an incentive for correctly reporting phishing and received it if they did so; or no special setting, as a control group.
\end{itemize}

To form the final experimental groups, we made deterrents and incentives mutually exclusive. 
Therefore, employees were assigned to one of the three different settings related to timing (immediate, delayed, and no training), and one of the three different settings related to incentives and deterrents (deterrent, incentive with reward, and no special setting).
The cartesian product between these settings generated $\mathbf{3 \times 3 = 9}$ experimental groups of similar sizes: from \smallerGroup participants in the smallest to \largerGroup participants in the largest. We display these combinations in Table~\ref{tab:groups}, where we also report the sizes of each group, and how many participants in each group fell for at least one of our phishing tests.
Crucially, the control group that did not receive any form of training, deterrent, or incentive was not informed if they failed a phishing test---this is necessary to assess the magnitude of the reminder component of the exercise.

\subsection{Study Execution}
\label{sec:expsetup.execution}

The study lasted from August to October 2021.
During this time span, participants received 3 different simulated phishing emails in random order and at random time intervals, with at least two weeks in between consecutive emails. 
For this, the study was divided into three 2-weeks windows where participants could receive their email. Emails were sent 7am-4pm, Monday to Friday.
We picked this interval such that any potential nudging effect is likely to happen, as retention from phishing exercises was measured to be around 10-30 days~\cite{jampen2020don}; the frequency of emails was slightly higher than the usual frequency of the company, which has high variance between 2 and 6 weeks.

Participants were unaware that they were part of the study and would receive phishing tests, to avoid modifying their behavior. This is common practice in such studies~\cite{jampen2020don,marshall2023exploring}---we discuss its ethical implications further in Section~\ref{sec:expsetup.ethics}.

We designed the 3 simulated phishing emails (provided in Appendix~\ref{sec:extra.emails}) in collaboration with the partner company and by leveraging common elements of phishing~\cite{zhuo2023sok}, such as creating a sense of urgency and prospecting consequences for inaction.
All three emails were tailored to the company, simulating broad phishing campaigns targeting the whole organization, but they were not spearphishing: for example, they did not contain any data regarding the recipient nor tried to impersonate any real employee or manager at the company. 
Our company operates in four different languages, therefore all emails were translated to the recipient's work language out of these four---an information an hypothetical attacker could infer from, e.g., the recipient's name or location of their branch.

Emails contained a link to a phishing website where participants were asked to enter their login credentials.
For this reason, the company recorded both \textbf{clicks} on the link contained in the simulated emails and further \textbf{dangerous actions}, i.e., submitting their credentials to the phishing website.
We considered a test failed only when participants fell for the phish: while clicking on phishing links might not be desirable from the company's standpoint (and thus we also record and analyze clicks), it is a common behavior to inspect the website before deciding whether it is legitimate. This treatment of clicks and dangerous actions is similar to previous works on phishing~\cite{marshall2023exploring}.
Neither we nor the partner company registered the data input by users on our simulated phishing websites; in particular, this means we do not know whether participants entered their real credentials or bogus ones.
This means that our recorded dangerous actions might be overestimated, however, this trade-off ensures maximal safety for participants.
The company also recorded employees' reports of phishing emails.

At the end of the experiment, participants were debriefed via an internal email by the partner organization's IT security, which allowed opt-out: 14 did so and were excluded from the our study.
Then, we conducted semi-structured interviews with the aforementioned participants who accepted our invite at the end of the study.
These happened online over a videoconferencing platform provided by the organization.
Every interview started by reading a consent text approved by our IRB, and obtaining explicit consent from the participant. Both the email invitation and this text reassured the confidentiality of all answers and that the interview was conducted by a researcher; yet, we remark that the interviewed participants are voluntaries.

\subsection{Ethics and Participants' Safety}
\label{sec:expsetup.ethics}

The study was approved by our institution's Institutional Review Board (IRB) and conducted in collaboration with the CISO and Data Protection Officer of the partner's company.
In the following, we discuss the most critical points regarding our study and how we ensured participants' safety while still conducting an ecologically valid experiment.

\paragraph{Use of Deception}
While the employees of the company are generally aware that they can receive simulated phishing emails, participants were not informed of our study in advance, to avoid changing their behavior~\cite{parsons2015design}.
For this, we received a waiver of informed consent.

To justify the use of deception, we followed the guidelines from previous studies on phishing~\cite{parsons2015design}.
First, this study could not have happened otherwise with the same ecological validity: it is a ``natural observational'' study, in which data is collected as it occurs in the real world. We posit that study participants made aware of being observed or warned of potential deception in advance will adjust their behavior.
Second, a waiver must not adversely affect the welfare of subjects: we argue this applies to our study because participants already know they might receive deceptive emails from the company. Further, there are no consequences for study participants if they fail phishing tests----besides participants in our \textit{deterrent} group, whose consequence is to undergo mandatory training on phishing after failing two tests, which does not have negative consequences.
Lastly, participants need to receive a complete debriefing and a means to opt out of the study: we sent a detailed explanation of our experiment as a debriefing email at the end of the study, and participants had 60 days to reply and opt out if they wished.

\paragraph{Participants' Safety}
The main source of distress for participants is the reception of simulated phishing emails, which might interrupt their workflow; further, falling for one of our phishing tests might create discomfort and embarrassment.
We reason that our study does not contribute to increasing the stress of participants, as phishing tests are a well-established practice in our partner company, and employees are aware of it.
Our simulations also do not carry the same negative consequences of falling for a real phishing email, which participants might still receive on their day-to-day jobs.
Finally, negative feelings as a result of our study are possible, e.g., loss of trust in the company, or a false sense of security in case the training is misunderstood and participants think they were protected by real phishing. We argue that these risks are not directly caused by our study but rather by the widespread industry practice of phishing simulations that our study aims to shed light on.
We further mitigate this risk by ensuring that there are no repercussions for participants failing a phishing test besides being potentially required to undergo training on email safety.
Finally, semi-structured interviews happened via videoconferencing, due to the COVID-19 pandemic.

\paragraph{Data Anonymity}
All experimental data was collected by the partner company, and was provided to us in pseudonymous form.
Every participant was identified by a random identifier that the company generated, and that linked together relevant Personally Identifiable Information (such as age, gender, and department) with participants' performance in phishing tests and suspected phishing emails they reported.
Further, we recall that we do not register data entered on the simulated phishing websites, nor whether the entered credentials were real or bogus.
Reported emails did not carry any PII: the data we received only recorded whether a reported email was one of our simulations, and the verdict by the existing anti-phishing system employed by the company; it did not contain any information regarding the email's subject, content, sender, and further recipients besides the participant.
Participants volunteering to get interviewed by us consented to disclose to us their name and their performance during the study. They were assured that their answers would be treated confidentially and not shared with their employer.

\section{Training Effectiveness}
\label{sec:results_effectiveness}
\begin{table*}[tb]
	\renewcommand{\arraystretch}{1}
	\centering
	\caption{Experimental groups we compare to study training effectiveness. We report the number of \textit{relevant participants} that failed at least one phishing test for each group, and the percentage of such participants that further clicked, respectively, performed the dangerous action, on two and three simulated phishing emails.}
	\rowcolors{2}{white}{gray!10}
	\begin{tabularx}{\linewidth}{@{}XXXX|XX|XX@{}}
		\toprule
		\rowcolor{gray!10}
		& & & & \multicolumn{2}{c|}{\textbf{\# of Clicks}} & \multicolumn{2}{c}{\textbf{\# of Dangerous Actions}} \\
		Group & Training type & Incentive /\newline Deterrent & Relevant\newline Participants & 2 & 3 & 2 & 3 \\
		\midrule

2 & \cellcolor{white} & Deterrent & 143 / 522 & 31.5\% & 5.6\% & 13.3\% & 0.7\% \\ 
3 & \multirow{-2}{*}{\cellcolor{white}Immediate} & No & 151 / 500 & 33.1\% & 12.6\% & 21.2\% & 0.7\% \\
\midrule
8 & \cellcolor{gray!10} & Deterrent & 136 / 529 & 26.5\% & 6.6\% & 14.0\% & 1.5\% \\ 
\textit{Control} & \multirow{-2}{*}{\cellcolor{gray!10}None} & No & 127 / 510 & 37.0\% & 15.0\% & 23.6\% & 7.9\% \\
		
		\bottomrule
	\end{tabularx}
	\label{tab:clicks_falls}
\end{table*}

We now analyze the results of our experiment to answer our first research question: which elements of embedded phishing training lead to improvements in employees' future phishing detection.

\paragraph{Methods}
We employ the commonly used metric of employee performance in detecting our simulated phishing emails, namely how many \textit{dangerous actions} we recorded for different groups.
As we are only interested in participants who would need phishing training, we exclude all who did not fail any phishing test and thus never received either training or deterrents if they were in those experimental groups.
Similarly, we exclude participants who only failed their last simulated email, because they would not have had the chance to demonstrate any improvement due to our deployed mechanisms during our study.
We report the number of these remaining participants for the groups that are relevant to our hypotheses in Table~\ref{tab:clicks_falls}, and the percentage of these participants that further clicked on a second or all three simulated phishing emails.

To analyze the effect of our deployed mechanisms, we first analyze our data for normality with a Shapiro-Wilk test. As our data is not normally distributed, we compare the performance of participants in the different groups with a Kruskal-Wallis test, followed by a post-hoc Dunn's test.
We consider a statistical test significant if the p-value is below $0.05$.

\subsection{Results}

\paragraph{H1: Embedded phishing training improves employees' future phishing detection}
We first study in isolation the effect of embedded training by comparing the performance of experimental group 3, who only received embedded training but no incentive or deterrent, with our control group.

\begin{figure}[tb]
	\centering
	\includegraphics[width=\linewidth]{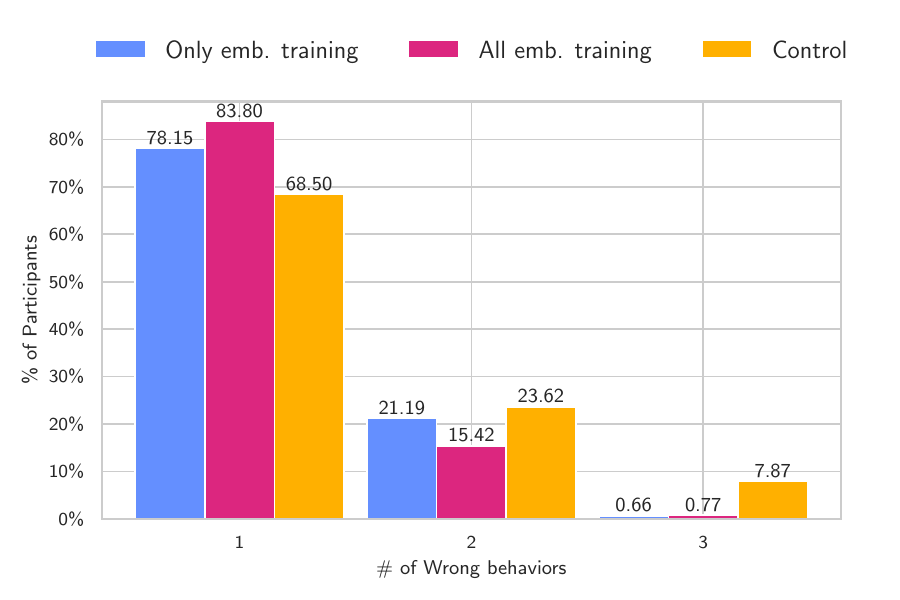}
	\caption{Dangerous actions for the embedded training groups. We show group 3 who only received embedded training, all groups receiving embedded training regardless of other conditions, and the control group.}
	\label{fig:training_perc}
\end{figure}

The performance of employees in the two groups is reported in Table~\ref{tab:clicks_falls} and in Figure~\ref{fig:training_perc}, where we display how many of all the participants failing one phishing test further fell for two or all three simulated emails.
We analyze the difference in the percentages of these ``recidivist'' participants between the two groups, to see whether the training served after the first mistake helped participants avoid further mistakes~\cite{lain2022phishing,caputo2013going}.
Interestingly, the difference between the two groups is not statistically significant regarding how many clicked on how many phishing emails: both groups performed similarly.
However, it is significant regarding the dangerous action ($H(1) = 4.36, p < 0.05$): while a similar percentage of employees fell for two simulated emails ($21.2\%$ in group 3 compared to $23.6\%$ in the control group), the most pronounced difference is in the number of employees falling for all three emails ($0.7\%$ compared to $7.9\%$).

\paragraph{H2: Deterrents improve employees' future phishing detection}
We then study the effect of deterrents by comparing the performance of experimental group 8, who only received deterrents but no training, with our control group. Recall that these deterrents were implemented as emails sent to participants after they failed a simulated email, thus inherently providing a nudge to avoid further mistakes.

\begin{figure}[tb]
	\centering
	\includegraphics[width=\linewidth]{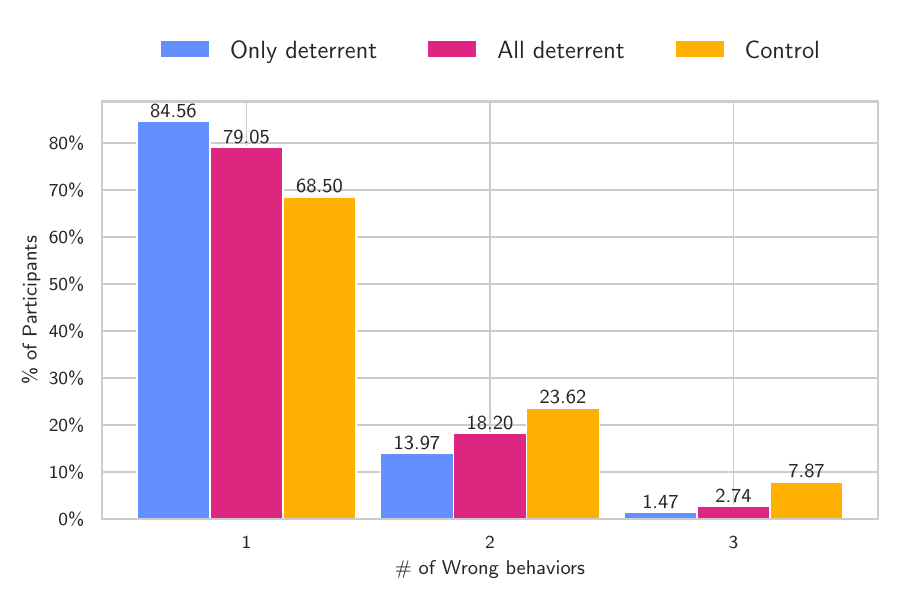}
	\caption{Dangerous actions for the deterrent groups. We show group 8 who only received our deterrent, all groups receiving deterrents regardless of other conditions, and the control group.}
	\label{fig:deterrent_perc}
\end{figure}

The difference between the two groups is statistically significant both for the number of clicks ($H(1) = 10.53, p < 0.05$) and for performing the dangerous action ($FH(1) = 10.26, p < 0.05$).
As reported in Table~\ref{tab:clicks_falls} and shown in Figure~\ref{fig:deterrent_perc}, the percentage of employees both clicking on the phishing emails and performing the dangerous action again is lower in the deterrent group compared to the control group: $26.5\%$ compared to $37\%$ for clicking twice and $6.6\%$ compared to $15\%$ for clicking all three times. Similar trends are observed for performing the dangerous action---$14\%$ compared to $23.6\%$ for two times and $1.5\%$ compared to $7.9\%$ for three times.
Therefore, our deployed deterrents seem effective in reducing the likelihood of employees falling for phishing emails.

\paragraph{H3: Embedded phishing training is not more effective than deterrents in improving employees' future phishing detection}
We now compare the two settings to understand which elements seem to provide benefits to employees.
We first compare the performance of group 3 (embedded training without deterrents) with group 8 (deterrents without training).
We can observe from the numbers reported in Table~\ref{tab:clicks_falls} that participants in the deterrent group clicked on the phishing emails less than participants in the training group, especially in clicking on all three emails ($12.6\%$ versus $6.6\%$): this is indeed statistically significant ($H(1) = 5.36, p < 0.05$).
Instead, the two groups perform similarly in terms of performing the dangerous action: their difference is not statistically significant.

\begin{figure}[tb]
	\centering
	\includegraphics[width=\linewidth]{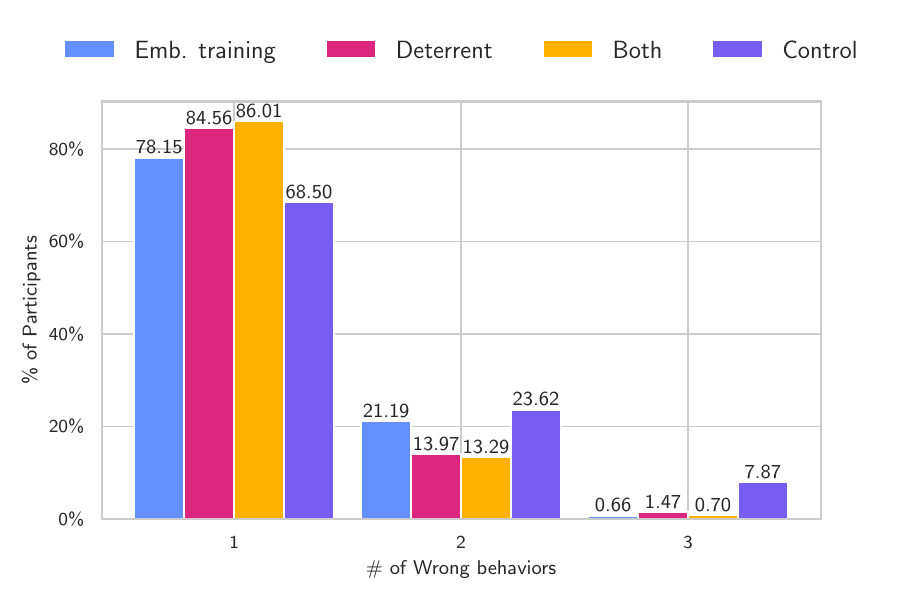}
	\caption{Dangerous actions for the embedded training and deterrent groups. We show group 3 who only received embedded training, group 8 who only received our deterrent, group 2 who received both, and the control group.}
	\label{fig:training_or_deterrent_perc}
\end{figure}

We also show the performance of participants in the two groups in Figure~\ref{fig:training_or_deterrent_perc}, where we also depict both the performance of the control group and of group 2, who received both training and deterrents.
We can observe that the performance of all three groups is similar.
Group 2, receiving both settings, presents a similar performance as the deterrent group 8: it is statistically significantly better than the control group both in clicking on the phishing emails again ($H(1) = 8.15, p < 0.05$) and in performing the dangerous action again ($H(1) = 12.99, p < 0.05$).
However, it is not different from group 3 (only embedded training) or group 8 (only deterrents) in any of the two metrics.

We conclude that the usefulness of the two mechanisms is comparable and that the content provided by training does not seem to provide additional benefits to employees compared to the pure nudging effect of our deterrent emails.

\paragraph{H4: Followup mandatory training is not better than voluntary and no training for more susceptible participants}
Recall that we are using the actual deterrent to enforce training, i.e., participants in the deterrent group that failed twice are required to go through mandatory training.
This way, we can make sure that some participants actually consumed the training material, compared to the other groups where undergoing training is voluntary.
However, compared to previous analyses, here we must handle participants more carefully: participants receiving mandatory training are the most susceptible ones of their group, failing two tests.
Therefore, we only consider participants who failed two or three simulated emails in both group 8 and the control group in the following analysis to compare similar populations.

We ran a Kruskal-Wallis test and found that there is no statistically significant difference between participants receiving mandatory training and participants in the control group; i.e., no difference between the two groups when comparing who failed twice (and potentially received mandatory training) and who failed all the times.
Similarly, we observe a non-significant difference for this population between participants undergoing mandatory training and similar participants receiving voluntary training (group 3).
Therefore, we conclude that for the most susceptible participants, mandatory training did not provide additional benefits.

\section{Training Improvements}
\label{sec:results_improvements}
We now analyze the results of the experiment to answer our research question relative to improvements to the process of phishing training: a new form of \textit{delayed} training and positive incentives for reporting phishing in the form of a reward.

\paragraph{Methods}
To test our hypothesis related to delayed training, we again only consider participants who failed at least one phishing test because only they were then exposed to the training material.
To test our hypotheses related to rewards, instead, we consider all participants in group 9, because everyone was exposed to the email informing them of the reward at the beginning of the study.

We then apply the same methodology detailed in Section~\ref{sec:results_effectiveness}, with the addition that we now also consider how many simulated emails participants correctly reported, and how many overall.

\subsection{Results}

\begin{table*}[tb]
	\renewcommand{\arraystretch}{1}
	\centering
	\caption{Experimental groups we compare to study delayed training. We report the number of \textit{relevant participants} that failed at least one phishing test for each group and the percentage of such participants that further clicked, respectively, performed the dangerous action for two and three simulated phishing emails.}
	\rowcolors{2}{white}{gray!10}
	\begin{tabularx}{\linewidth}{@{}XXXX|XX|XX@{}}
		\toprule
		\rowcolor{gray!10}
		& & & & \multicolumn{2}{c|}{\textbf{\# of Clicks}} & \multicolumn{2}{c}{\textbf{\# of Dangerous Actions}} \\
		Group & Training type & Incentive /\newline Deterrent & Relevant\newline Participants & 2 & 3 & 2 & 3 \\
		\midrule

		1 & \cellcolor{white} & Incentive & 110 / 483 & 39.1\% & 3.6\% & 17.3\% & 0.0\% \\ 
2 & \cellcolor{white} & Deterrent & 143 / 522 & 31.5\% & 5.6\% & 13.3\% & 0.7\% \\ 
3 & \multirow{-3}{*}{\cellcolor{white}Immediate} & No & 151 / 500 & 33.1\% & 12.6\% & 21.2\% & 0.7\% \\
\midrule
4 & \cellcolor{gray!10} & Incentive & 103 / 480 & 37.9\% & 6.8\% & 20.4\% & 3.9\% \\ 
5 & \cellcolor{gray!10} & Deterrent & 135 / 516 & 28.1\% & 8.2\% & 20.0\% & 0.0\% \\ 
6 & \multirow{-3}{*}{\cellcolor{gray!10}Delayed} & No & 146 / 534 & 31.5\% & 5.5\% & 13.7\% & 0.0\% \\
\midrule
\textit{Control} & \multirow{-1}{*}{None} & No & 127 / 510 & 37.0\% & 15.0\% & 23.6\% & 7.9\% \\
		
		\bottomrule
	\end{tabularx}
	\label{tab:clicks_falls_delayed}
\end{table*}

\paragraph{H5: Embedded and delayed training seem equally useful}
We start by analyzing the effectiveness of delayed training by comparing the performance of group 6, who only received such training, with the control group. We report the percentages of participants clicking and falling for two and three phishing emails in Table~\ref{tab:clicks_falls_delayed}.
We find that delayed training proves more effective both in reducing the number of clicks ($H(1) = 8.09, p < 0.05$) and the number of dangerous actions ($H(1) = 13.94, p < 0.05$) than control.

\begin{figure}[t]
	\centering
	\includegraphics[width=\linewidth]{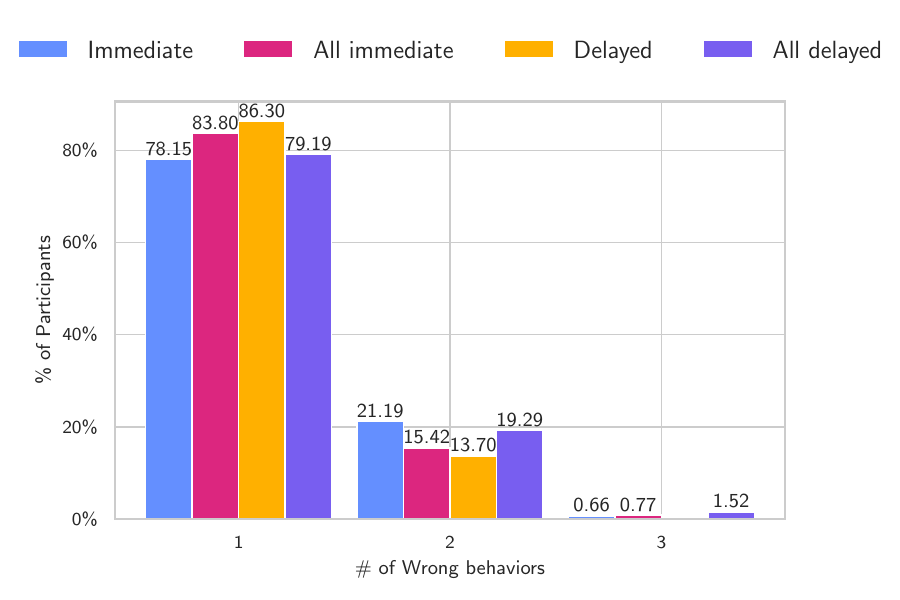}
	\caption{Dangerous action by type of training. We show group 3, who only received immediate training, all other groups receiving immediate training, group 6, who only received delayed training, and all other groups receiving delayed training.}
	\label{fig:training_type_perc}
\end{figure}

We then compare the performance of the two types of training by comparing group 6 with group 3, who only received immediate training. We find that they perform equally well, and that their difference in performance regarding both clicking and performing the dangerous action is not significant.
This is also shown in Figure~\ref{fig:training_type_perc}, where we show the percentage of participants that clicked and performed the dangerous action for these two groups, and for all the remaining groups that received immediate and delayed training.

\begin{figure}[t]
	\centering
	\includegraphics[width=\linewidth]{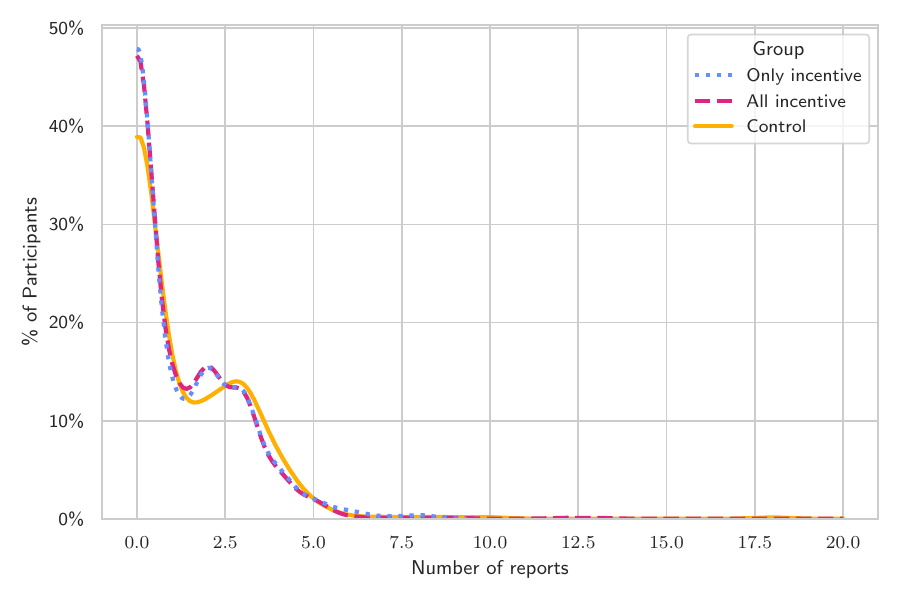}
	\caption{Distribution of participants by number of reports. We show group 7, who received rewards for correctly reporting phishing, all other groups receiving such rewards, and the control group.}
	\label{fig:reported_emails}
\end{figure}

\paragraph{H6: Rewards do not increase the rate of reported emails}
We compare how many emails were reported by participants in group 7, who were informed they would receive a reward if they correctly reported a phishing email, compared to the report group.
We recall that for this analysis we consider the whole groups, and not only participants who failed at least one phishing test.
We depict the percentage of employees that reported a given number of emails during our study in Figure~\ref{fig:reported_emails}.
We find that the rate of reported emails is not significantly different between the two groups.
We also find no difference in the increase of reported emails if we consider how many each participant reported in a time window before the experiment of the same length as the experiment duration.
The same applies if we extend our analysis to the other groups receiving incentives (groups 1 and 4).

Similarly, incentives do not seem to improve the accuracy of identifying phishing emails---a metric that is also anyway often secondary in organizations that prefer employees being overly cautious and report when in doubt.

\paragraph{H7: Rewards do not improve future phishing detection}
Similarly to the rate of reported emails, we also find no signficant difference in correctly identifying phishing emails when comparing group 7 and the control group.
We therefore observe that this setting in isolation does not seem to invite employees to pay more attention to phishing emails.

\section{Training Reception}
\label{sec:results_reception}
Finally, we now study the results of our experiments from the perspective of participants' perception of the training practice.

\paragraph{Methods}
 we interviewed 25 participants at the end of our study, asking them about their experience with the training exercises.
We recall from Section~\ref{sec:expsetup.participants} that these participants are volunteers among a pool of 200 invited ones.
The format of the interview was semi-structured, with open-ended questions that allowed participants to express their thoughts freely.
These questions touched on the main points we wanted to investigate; we report the script that the interviewer followed in Appendix~\ref{sec:extra.script}.

Participants were assured that the interviewer was a researcher conducting a study and that their answers would not be reported to their employer.
The interviews were conducted in the native language of the interviewee and interviewer.
We recorded the interviews, transcribed them, and translated them into English.
The interviews were then coded~\cite{deterding2021flexible} by one of the authors, with a deductive/top down approach; we based our codes on our research questions and the main points of the interview script (see Appendix~\ref{sec:extra.script}): (i) remembering the phishing exercises and training page; (ii) whether content was reviewed and what they remembered; (iii) what they found most and least beneficial; (iv) how they felt about the practice; (v) their mental image and understanding of phishing training; and (vi) their perception of the reward for reporting phishing.
We report the full coding scheme in Appendix~\ref{sec:extra.codebook}.

\paragraph{Most interviewed participants remember seeing the training page}
Almost all interviewed participants who made two or more mistakes, thus being exposed to the training page, remember seeing it and that it appeared after they entered their credentials.
The memory of the mistake they made and that the page appeared afterward to inform them also seems to be clear for most participants
\qt{13}{Yes, I saw it once recently when I clicked on the link - I have seen it somewhere before yes}.

\paragraph{Most interviewed participants did not go through the content}
Despite most participants remembering the training page, almost all participants explicitly stated they did not go through the content.
While this confirms the empirical measurements of recent work~\cite{sutter2022avoiding}, our work sheds further light on the reasons: several participants report lack of time
\qt{13}{That was then when I quickly went through the many emails during my holidays. Because of the time I didn't take a closer look at the page}.
The other main reported reason is that it is clear to participants that they failed a phishing exercise and do not need to know more
\qt{26}{I just didn't pay attention and didn't go any further down. I closed it as soon as I knew it was a phishing email test}.
The nudge and reminder of awareness are what is remembered the most from the exercise itself, as the first part of the training page often prompted participants to close the page
\qt{15}{The second time it was enough when I saw the fish [an image in the training page], I just thought 'Ok, thanks a lot'}
or
\qt{20}{I'm not sure if there was a lot of information, I thought it was just that I fell for it}.
These insights represent new findings on the reasons why training content yields mixed results in improving employees' behavior~\cite{jampen2020don}.

\paragraph{Susceptibility of interviewed participants comes from lack of attention rather than lack of knowledge}
The few participants who did go through the content of the training page felt they already knew the information on it
\qt{5}{And then it listed the 'golden rules' on how to recognize a phishing email: insecure websites, no HTTPs, sender, etc,.. all the rules you know}.
This is despite these participants falling for all the phishing emails in the study.
This supports the findings of a previous small-scale study on 3 ``repeat clickers'' that also observed them being knowledgeable on phishing cues~\cite{canham2023repeat}; our findings are significant because the common perception of embedded training is that it improves knowledge~\cite{zhuo2023sok}.

Instead, our study results suggest that that these employees do not lack the knowledge, but struggle anyway in paying attention: one of the main reported reasons is the pressure of work and the number of emails they have to process
\qt{10}{It's the mass that kept from looking closer. You just click: Next, next, next, thinking you should actually move forward};
some do not even perceive the need to pay attention
\qt{22}{[Training] didn't help in the sense that I visited the link on all 3 because I just saw: ah, mail? Ah, link? That's just my way of working}.
All these participants indeed showcased good knowledge of what is phishing and its common cues; however, phishing is not always in their mind
\qt{11}{You look at more carefully but not in the way that you look at these already with the thought in mind that it could be phishing}
potentially also due to information overload
\qt{18}{You get so much information every day [...] it almost overwhelms you}.
This finding is relevant because it corroborates previous studies that concluded this only by measurements~\cite{williams2018exploring} or in artificial settings~\cite{sarno2022so,zhuo2023sok} but not qualitatively.

\paragraph{What interviewed participants found most beneficial is the periodic reminder to be vigilant}
As participants did not find the content of the training useful, we asked them which aspects of the practice they found beneficial.
Most appreciate it for its periodic reminder effect
\qt{10}{It's just important that you deal with it again in between times}
with many participants explicitly stating that they need such periodic nudges to keep their awareness high
\qt{5}{The effectiveness lies in the repetition, that you have to remind yourself again and again and do such actions. Even with someone like me, who would actually know that you have to look at it}.
Another appreciated element is that these exercises allow them to test their knowledge and make mistakes in a safe environment
\qt{20}{Of course you can do internal training or e-learning, but that's quickly forgotten and the best way to learn is to make a mistake yourself, and it's better to make it with an internal email}.

One interpretation of these findings is that training is not helping consolidate employees' knowledge about phishing, thus the need for periodic nudges, and a possible reason why previous studies find wide knowledge retention ranges~\cite{marshall2023exploring}.
Our observation in the domain of phishing resembles an effect studied in the larger domain of cybersecurity awareness---that it does not lead to long-term improvements in secure behaviors~\cite{bada2019cyber,van2020if}.

On a smaller scale, a few participants appreciate the awareness element of the exercises reminding them of the threat and making them more aware of its consequences
\qt{7}{Especially if you can see the damage that can really occur, I think you're more likely to say `Okay, we have to be able to avoid this at all costs'}.

\paragraph{Interviewed participants have generally a positive feeling towards the training practice}
The practice of training seems to be generally well received among our interviewed participants, with many explicitly mentioning they think it is a good idea to run such exercises
\qt{13}{I don't perceive it at all as being intrusive and I can understand why they want to sensitize us}.
Some also appreciate the relief coming from failing an exercise and not making a mistake in real life
\qt{27}{I'm also glad at this point that it happened to me in that context and not somewhere else.}
and that the mistakes in the tests made them more mindful, which they perceive as positive
\qt{22}{I don't get upset. I take it sporty}.
A minority of the participants expressed more negative sentiments towards the exercise and their mistakes
\qt{8}{Personally, you get upset about it. But you will look a bit more closely at the next email}.

Our observations seem to partially contradict previous studies where employees disliked the introduction of exercises~\cite{caputo2013going,volkamer2020analysing} which strained the relationship with management~\cite{kirlappos2015fixing,hielscher2023employees,volkamer2020analysing}. However, most of the interviewed participants seem highly aware that the company routinely employs phishing exercises that have been in place for years, potentially improving their acceptance of the practice.
Further, our volunteer population might be biased towards expressing more positive views either due to more interest in the topic or because they feel under ``scrutiny''---we discuss this further in Section~\ref{sec:discussion}.

Whether this training should be enforced was more controversial, with many opposing views, from those that think that training should even be more frequent 
\qt{10}{I think we really need more of these test emails that you send. Then you can really see}
to those that strongly oppose this idea as it would add too much to their workload
\qt{11}{We have so many appointments and other training courses that we have to attend and now that too. This is an additional burden},
with a few more intolerant participants even stating they would reduce the frequency of the exercises to monthly or quarterly.
Our findings support previous studies that observe different reactions to enforced training~\cite{williams2018exploring,caputo2013going} and its contribution to cyber fatigue~\cite{reeves2021encouraging}.

\paragraph{Interviewed participants have a mixed understanding of the nature of the exercises}
A potential source of ineffectiveness of training could be that employees do not understand the nature of the exercises~\cite{lain2022phishing}.
We find that this can be the case for some participants: while most showed awareness that the exercises are part of training practice, some thought at least one email was part of a real attack, and that the training page was part of a protection mechanism:
\qt{25}{I thought that was a specific feature that was developed by IT to protect us. But that's not the case}.

What our study also finds is a different type of overconfidence, also only hypothesized in previous work~\cite{lain2022phishing}: participants perceive they very rarely receive phishing at work compared to their personal addresses, the majority they see are tests, and those can be failed with little repercussions, as expressed by a participant who is confident they never encountered phishing in their workplace, despite the fact that they fell for two of our three test emails
\qt{4}{I've never received a real one, but they do test us, so to speak. And I've already failed twice, really, they did very well}
while others express little concern about the consequences of failing the exercises
\qt{8}{I was actually 100\% sure with both emails that it was a bad email, but I opened it anyway}.

\paragraph{Incentives to secure behaviors are seen as unnecessary}
The recall of our reward was mixed, with only a few participants being aware of it either through our initial email or through seeing the box of chocolate being posted on social media by their colleagues.
Further, the general sentiment towards the reward was not positive, with only a few participants appreciating the idea, while almost all stated they would not get influenced by such measures
\qt{13}{I report the ones I see because I have to, it doesn't bother me if it comes with chocolate, but it's not a motivation to do more}
or even actively opposed the idea, associating reporting phishing with one of their duties
\qt{23}{I don't think you have to reward people for doing the right thing}.
This aligns with our quantitative results that observed no effect of the introduced rewards on phishing detection or reporting.

\section{Discussion}
\label{sec:discussion}
We now further discuss our findings. First, we summarize our results and how they help explain recent literature on phishing training. 
We then discuss practical advice and improvements to the practice that can be derived from our findings. 
Finally, we discuss the limitations of our study.

\subsection{Our Findings and Prior Literature}
Our results all suggest that the content component of embedded phishing training is not responsible for significant improvements in employees' vigilance: training did not perform better than simple threats of repercussions, participants did not find its content useful or even perceived they had to read it.
Comparing what is in common between our tested training and deterrents, we find it is its reminder nature: this was further confirmed by interviewed participants who felt more vigilant after failing a phishing test.
This finding offers us a way of reading recent results that found improvements after training despite participants not consuming its content~\cite{jansson2013phishing,sutter2022avoiding}, as these improvements are potentially caused by this reminder nature.
It also offers a new perspective on studies measuring the retention of phishing training~\cite{kumaraguru2010teaching,kumaraguru2007getting} by suggesting that what fades is the participants' alertness. 
We further observe that studies finding no improvements from training~\cite{lain2022phishing,caputo2013going} employed emails that were more spaced apart than ours and than the aforementioned retention studies~\cite{kumaraguru2010teaching,kumaraguru2007getting}, thus potentially too far apart to measure this effect.

\subsection{Improvements to the Practice}
From our findings, we therefore raise the question of whether the practice should be rethought and split into its components: it is difficult to replace phishing exercises as a metric for CISOs~\cite{hielscher2023employees} and element of security compliance, but its knowledge components should be moved to dedicated sessions, instead of delivering well-known ``security bits'' that get ignored in the context of phishing tests.
As the benefit seems to be in the reminder, we ask ourselves how the practice should evolve around this and what is the best way to deliver such reminders for alertness instead of the current ``gotcha'' approach of deceptive tests.
This further recommends fostering a security culture in organizations, for example, encouraging to report when in doubt~\cite{marin2023influence,lain2022phishing} or change phishing tests to ``drills'' with emails announcing themselves as test phishing email containing instructions and tasks to perform such as reporting it~\cite{linton2024fire}, akin to fire drills who evolved from surprise tests to scheduled exercises.

\subsection{Limitations}

\paragraph{Study design}
Due to our study design, we did not study nudges in isolation---they were always associated either with the training material, i.e., the content appearing after a mistake, or with the threat of repercussions, which might have influenced our participants' perception and attention to emails.

We could not further test the impact of enforcing training for a larger population, but only for the most susceptible participants. This was a limitation set by our partner company.
However, we believe that our qualitative insights on the perception of training content by the employees mitigate this limitation.

\paragraph{Length of study}
Our study is limited in time to 6 weeks and 3 simulated phishing emails.
Therefore, the lack of further improvements in employees' performance could be due to the short duration of the study, as the training might need more time to be effective.
However, our study lies in the framework of phishing training and education that has been in place for many years in the company---therefore, even comparisons on a short time frame are relevant.
Indeed, the short timeframe during which participants in the control group did not receive embedded training was enough to manifest differences in performance: we believe this is further proof of the reminder-only nature of the exercise and that the length of our study suffices.
In our study, we did not consider training frequency and defer it to future work.

\paragraph{COVID-19}
Our study happened during the COVID-19 pandemic, which is known to have amplified phishing attacks~\cite{bitaab2020scam} due to the increase in remote work, general confusion caused by the pandemic (e.g., rapid changes of corporate processes), and increased reliance on digital communication~\cite{defilippis2022impact}.
However our study happened at the end of the ``second lockdown'' in our country, as employees were returning to the office in the two months preceding our study.
Further, a good part of our heterogeneous participants who work in the field did not see significant changes in their work.
Therefore, while this might have influenced our participants' performance (e.g., our email concerning renewing one's badge access rights was the most successful), we believe that the general trends we observed are not influenced by the pandemic.

\paragraph{Simulated emails}
The simulated emails we used had varying performances, i.e., the average click and fall rates for the three emails were 33\%--21\%, 43\%-38\%, and 73\%-63\%. While one email outperforming the others could have biased our comparisons of improvements in employees' performance, we randomized the order of the emails to mitigate this effect.
The numbers we recorded for dangerous actions could be overestimated, as we did not record whether participants entered their real credentials on the phishing websites or bogus ones.

\paragraph{Word-of-mouth effect}
We cannot exclude colleagues noticing phishing emails and discussing them, potentially increasing their colleagues' vigilance and performance in the study.
However, previous studies found this effect to be limited~\cite{lain2022phishing}; moreover, our participants were a random sample of less than 10\% of the employee base, limiting the likelihood of immediate colleagues being enrolled in the study.
Finally, while we observed both click- and fall-rates slightly decreasing over the three sending windows (however, this effect is amplified by the effects of training and nudges improving participants' performance), one of the interview questions asked whether phishing exercises were discussed, and no participant reported this.

\paragraph{Interviewed participants}
The participants we interviewed were all volunteers, which might have led to a selection bias. They might have had more positive sentiments towards the training or been curious about it. 
Further, many of them fell for the phishing emails multiple times, which might have made them feel under scrutiny and potentially influenced their responses, despite our assurance that the interviews were anonymous and would not be shared with their employer.
Recall that we took several countermeasures to encourage participants to speak freely, such as the assurance of anonymity, privacy of their answers, and having researchers conduct the interviews---see Section~\ref{sec:expsetup.execution}.

\paragraph{Other confounds}
There are a few remaining confounding factors we did not measure in our study. 
Participants were mostly aware of the existence of phishing tests; some of the interviewed most susceptible participants seemed aware that there are generally no consequences for failing them, as they saw the training (and thus failed for previous simulations) before our study.
Further, the nature of our chosen reward might influence our results---it is possible that better incentives (e.g., monetary rewards) could have a different impact.
Finally, while we know that our participants were diverse in terms of roles, responsibilities, and seniority, we do not know what is their use of technology and the Internet in their job, which is known to influence performance in phishing tests~\cite{lain2022phishing,jampen2020don}.

\section{Conclusion and Future Work}
\label{sec:conclusion}
In this paper, we investigated several aspects of embedded phishing training in organizations by running a large-scale mixed-methods study in a partner company.
One common hypothesis about the effectiveness of phishing, supported by our interviews, is that the main problem is lack of attention rather than lack of knowledge~\cite{canham2023repeat,canham2019enduring,bada2019cyber,butavicius2022people}.
Indeed, our novel quantitative results show that nudges (in the context of deterrents) work as well as the training material.
This therefore raises the question of whether to reconsider this (expensive~\cite{brunken2023properly} and controversial~\cite{caputo2013going,lain2022phishing,hielscher2023employees}) industry practice that was built around training knowledge~\cite{kumaraguru2010teaching}, while reminders came as a side effect.
Our study further invites caution in employing this type of training due to the finding that it has the potential to create misunderstandings and overconfidence.
We ask ourselves whether the practice and its reception can therefore be improved by focusing on the component that actually works, i.e., the nudging effect.
For this, further research is needed to understand what is the optimal way to deliver such reminders, how frequent this should be (which we did not consider in this study), and whether nudges can be separated from threats of negative consequences.

We further uncovered that time pressure is a significant factor not only in phishing susceptibility, but is also responsible for employees not paying attention to the training content. Moreover, delaying the delivery of training is as effective in reducing the likelihood of failing future phishing tests as immediate training, which recommends that the timing of training should be further studied.

\bibliographystyle{ACM-Reference-Format}
\bibliography{bibliography}

\appendix
\section{Supplementary Material}
\label{sec:extra}
\subsection{Further Details on Population}

\begin{figure}[ht]
    \centering
    \includegraphics[width=\linewidth]{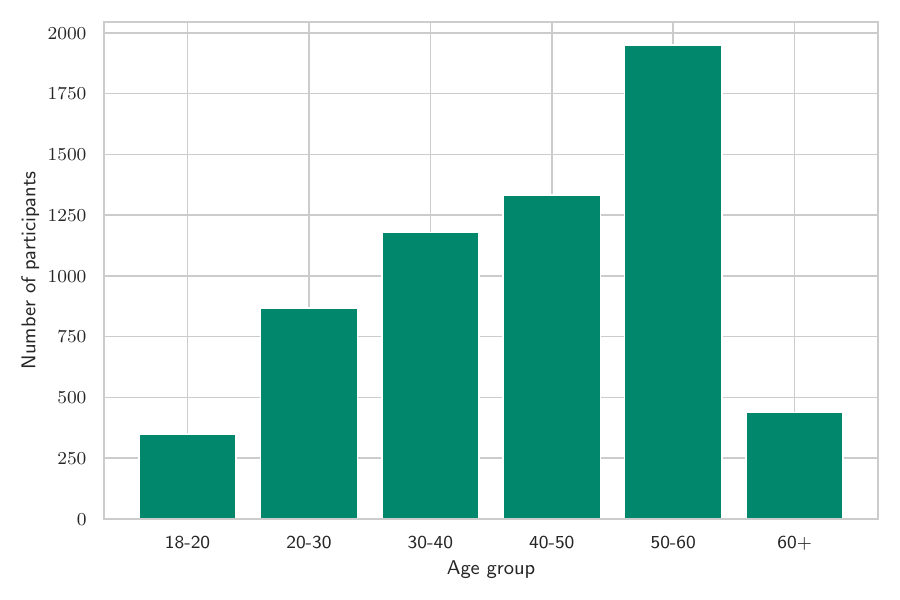}
    \caption{Age distribution of the participants in our study.}
    \label{fig:age_groups}
\end{figure}

We report in Figure~\ref{fig:age_groups} the age distribution of our participants. We observe that the population is skewed to long-term employees that are in the 50-59 age range, in line with the company's demographics.

\subsection{Phishing Emails}
\label{sec:extra.emails}

We report in Table~\ref{tab:emails} the English translation of the content of two of the three phishing emails we used in our study. We designed them to be realistic and to mimic the style of phishing emails that employees might receive in their work environment.
The third email is omitted due to being designed with the look and feel of a specific product that the company sells but also employs internally, which could reveal the company's identity.
We omit the senders of the emails: for Email 1, the sender was a typosquat of the company's name; for Email 2 and Email 3, the sender was an impersonating domain (e.g., \texttt{[company]-internal.com}).

\begin{table*}
    \centering
    \caption{Content of the two text-only simulated phishing emails used in our study. Text in square brackets represents placeholders for values, or redaction of the company's identity. We omit the sender, which for Email 1 was a typosquat of the company's name, and for Email 2 was an impersonating domain.}
    \label{tab:emails}
    \begin{tabularx}{\linewidth}{p{0.5\linewidth}|p{0.5\linewidth}}
        \textbf{Email 1} & \textbf{Email 2} \\
        \toprule

        \textbf{IMPORTANT: Register for new time tracking software}
        & 
        \textbf{[COMPANY] Badge: Renewal of your building access rights} \\
        & \\
        Dear Employee,\newline

        Earlier this year we announced the rollout of a new time tracking software ``TimeTrack'' across the whole company. The software will be used to track working time, vacation days and leave of absences. You are required to register with the new tool until [NEXT WEEK], please follow the link below:\newline

        [LINK]\newline

        We'd like to remind you that all requests for vacation days after the [NEXT WEEK] have to be resubmitted and approved. \newline

        Best, \newline
        Your TimeTrack-development team
        & 
        
        Dear Employee,\newline

        Since many of you have been working from home for a while due to the Coronavirus home office restrictions, we'd like to remind you that our security policy dictates that building access rights for your badges need to be renewed every 3 months. \newline

        Click here [LINK] to update your building access rights.\newline

        If your access rights are not updated, you may not be able to enter your workplace.\newline

        Best, \newline
        Your [COMPANY] IT Security Team

        \\
        \bottomrule
    \end{tabularx}
\end{table*}

\subsection{Study Emails}
\label{sec:extra.study_emails}

We report in Table~\ref{tab:study_emails} the content of the emails we employed in this study: (i) the initial email informing participants in the \textit{Incentive} groups of the reward for reporting phishing emails; (ii) the two emails sent to participants in the \textit{Deterrent} groups after the first and second mistake made on our phishing emails; and (iii) the email sent to participants in the \textit{Delayed} training groups informing them of the mistake they made and linking the training material.
All emails were signed at the end by the Information Security team of the company, who was responsible to send them.

\begin{table*}
    \centering
    \caption{Content of the emails we used in the study to inform participants in the \textit{Incentive}, \textit{Deterrent}, and \textit{Delayed} training groups. Text in square brackets represents placeholders for values, or redaction of the company's identity. All emails were signed at the end by the Information Security team of the company.}
    \label{tab:study_emails}
    {\footnotesize
    \begin{tabularx}{\linewidth}{p{0.1\linewidth}|X}
        \textbf{Email} & \textbf{Content} \\
        \toprule

        {\normalsize Incentive}
        & 
        \textbf{Rewards for reporting phishing emails}

        Dear [NAME],\newline

        Phishing emails represent a significant threat to the security of our IT infrastructure. Phishing emails refer to deceptive emails that trick users into performing unsafe actions, such as opening unknown attachments or entering user credentials. The ideal way to handle such emails is to report them using the ``Report Phishing'' button in Outlook: [URL]\newline

        To improve the security of our IT infrastructure by reducing successful phishing attempts, we are now introducing a pilot program where selected employees receive rewards for successfully identifying and reporting phishing emails. The reward will take the form of a small gift and is distributed at the end of an internal assessment cycle (roughly 3 weeks).\newline
        \\ \midrule

        {\normalsize Deterrent \newline (first mistake)}
        &
        \textbf{Oops! You clicked on a simulated phishing email}

        Dear [NAME],\newline

        Our system has determined that you performed a dangerous action in response to a simulated phishing email yesterday. These emails are part of [COMPANY] internal phishing training, and are therefore harmless. However, in general phishing emails represent a significant threat to the security of our IT infrastructure and all employees.\newline

        In an effort to reduce the risk of successful phishing attacks, repeatedly performing dangerous actions, such as opening unknown attachments and entering user credentials on fraudulent sites, will result in a mandatory phishing training for the employee. \textbf{The next time you perform a dangerous action in response to a phishing email, you will be required to attend a phishing e-learning.}\newline
        \\ \midrule

        {\normalsize Deterrent \newline (second \newline mistake)}
        &
        \textbf{Oops! You clicked on a simulated phishing email again}

        Dear [NAME],\newline

        Our system has determined that you performed a dangerous action in response to a simulated phishing email yesterday. These emails are part of [COMPANY] internal phishing training, and are therefore harmless. However, in general phishing emails represent a significant threat to the security of our IT infrastructure and all employees.\newline

        In an earlier email you were informed that repeatedly performing dangerous actions in response to a simulated phishing email will result in mandatory phishing training. \textbf{As a consequence you are required to complete the e-learning in the next week.} All information about and the training can be found here: [LINK].\newline
        \\ \midrule

        {\normalsize Delayed \newline training}
        &
        \textbf{Oops! You clicked on a simulated phishing email}

        Dear [NAME],\newline

        Our system has determined that you performed a dangerous action in response to a simulated phishing email yesterday. These emails are part of [COMPANY] internal phishing training, and are therefore harmless. However, in general phishing emails represent a significant threat to the security of our IT infrastructure and all employees.\newline

        To avoid falling for phishing emails in the future, please review the material provided by [COMPANY] on our phishing information page: [LINK].\newline
        \\\bottomrule
    \end{tabularx}
    }
\end{table*}

\subsection{Interview Script}
\label{sec:extra.script}

We report in the following the script we used as a guideline for our semi-structured interviews.
For brevity, we omit the researcher introducing themselves as a researcher who is conducting a study for our institution.
The script was adapted slightly for our ``expert'' participants who did not fail any phishing test and receive training.

\paragraph{Introduction}
{
\itshape
Over the last 2 months, you have received 3 simulated phishing emails as part
of this study. The aim of this study is to understand the behavior of [company]
employees in relation to phishing emails and to find ways to improve the
training and awareness of employees.
We interview [company] employees with all kinds of backgrounds, experiences
and positions who have clicked on none, several or all of the emails. You have
been invited to this interview because you clicked on at least 2 of the 3
emails.
We are not here to rate you and we are aware that everyone makes
mistakes. My colleague and I have also fallen for phishing emails and we
know that it's unpleasant and you may not like to talk about it. We are just
trying to understand why this happens and how the [company] can help its
employees to be better prepared and protected.

If this conversation becomes unpleasant, you can speak up at any time
and change the subject or break off.
If you agree, I will record this meeting and transcribe it later. All your
answers are confidential and will be stored anonymously. No individual
answer will be passed on to your employer. You have the right to
stop this interview at any time without explanation and to withdraw
from the study by informing me; your answers will then be deleted
immediately.
}

\paragraph{Background}
{\itshape
\begin{itemize}
    \item How often do you use your computer at work? How would you rate
    your knowledge of computer security and data protection?
    \item How familiar are you with the term "phishing" emails? Could you explain what it means? Have you ever come across such e-mails?
    \item What's the worst thing that can happen when you click on a phishing email?
\end{itemize}
}

\paragraph{Phishing Links and Attachments}
{\itshape
\begin{itemize}
    \item Did you see any phishing emails in the last 12 months in
    connection with your work? How did you handle the situation?
    \item When managing emails, do you actively think about phishing?
    \item Do you feel protected by the [company] against phishing emails?
    \item Did you notice that the email looked suspicious? If so, what made it
    suspicious to you?
    \item Have you reported the e-mail as phishing or similar?
    \item Did you click on the link? Why and what happened?
    \item Did you enter your credentials? What were you
    thinking?
\end{itemize}
}

\paragraph{Training Webpage}
{\itshape
\begin{itemize}
    \item Do you remember seeing this website in the last 12 months? In what
    context did you come across it?
    \item How did you feel when you saw the website?
    \item How do you think you ended up on this page? What happened?
    \item How much time do you estimate you spent on the training site? Did
    you study the content? If not, why? Did you find the content of the
    page useful?
    \item Did you think the site was trustworthy? Why/why not?
    \item Has your attitude towards suspicious emails changed after you came
    across this page?
\end{itemize}
}

\paragraph{Incentives and Deterrents}
{\itshape
\begin{itemize}
    \item Do you remember receiving information about any of these practices?
    Which one?
    \item In what context did you receive this information?
    \item Did you trust this information? If not, have you tried to confirm the
    information with IT security or your supervisor?
    \item How did you feel when you first heard about these practices?
    \item How has this information influenced your behavior when checking
    your e-mails?
    \item Has the information on these practices led to discussions on the
    subject? If so, with whom?
    \item Has your attitude towards your employer changed? If yes, how?
    \item Did you take any action when you found out about the introduction
    of these practices?
\end{itemize}
}

\subsection{Interviews Codebook}
\label{sec:extra.codebook}

We report the codebook we used to code the interviews in the following: we report the main topics we covered, and the codes used for each topic.
\begin{itemize}
    \item Topic: \textbf{Perception of phishing and training}
    \begin{itemize}
        \item Familiarity with phishing.
        \item Reasons for clicking on the phishing emails.
        \item Perception of the training practice.
        \item Perception of the incentive to report.
    \end{itemize}
    \item Topic: \textbf{Memory of the phishing emails and training.}
    \begin{itemize}
        \item Memory of phishing emails at work.
        \item Memory of falling for the phishing emails.
        \item Memory of seeing the training page.
        \item Understanding of training.
    \end{itemize}
    \item Topic: \textbf{Perception of training content.}
    \begin{itemize}
        \item Reasons for not reviewing the training material.
        \item Recalled training content.
        \item Usefulness of the training material or practice.
    \end{itemize}
\end{itemize}

\end{document}